\documentclass[12pt,thmsa]{article}
\usepackage{sw20lart}
\usepackage{epsfig}
\usepackage[english,spanish,german,activeacute]{babel}

\newcounter{iiicounter}
\def\biilist{\begin{list}{{\hss\it\roman{iiicounter})\/}}{\usecounter{iiicounter}\setlength{\topsep}{0pt}\setlength{\itemsep}{0pt}\setlength{\leftmargin}{0.8cm}\setlength{\labelwidth}{0.65cm}\setlength{\labelsep}{0.15cm}}}
\def\eiilist{\end{list}}

\def\`#1{\if#1i{\accent18 \i}\else{\accent18 #1}\fi}
\def\'#1{\if#1i{\accent19 \i}\else{\accent19 #1}\fi}

\input tcilatex
\input figmacros

\begin{document}
\baselineskip=16pt
\selectlanguage{english}

\title{LPM effect and dielectric suppression in air showers }
\author{{\large
\textrm{A.~N.~Cillis, H.~Fanchiotti, C.~A.~
Garc\'{\i}a~Canal
and S.~J.~Sciutto \vrule depth \baselineskip width
0pt}} \\
{\large \textit{Laboratorio de F\'{\i}sica Te\'orica}}\\
{\large \textit{Departamento de F\'{\i}sica}}\\
{\large \textit{Universidad Nacional de La Plata}}\\
{\large \textit{C. C. 67 - 1900 La Plata}}\\
{\large \textit{Argentina}}\\
{\large \textit{e-mail: cillis@venus.fisica.unlp.edu.ar}}}

\maketitle

\input pstricks 
\vfil

\begin{abstract}
An analysis of the influence of the Landau-Migdal-Pomeranchuck (LPM) effect
on the development of air showers initiated by astroparticles, is
presented. By means of computer simulations using algorithms that emulate
Migdal's theory, including also the so-called dielectric suppression, we
study the behaviour of the relevant observables in the case of ultra high
energy primaries. We find that the LPM effect can significantly modify the
development of high energy electromagnetic showers.
\end{abstract}

\vfil\eject

\section{Introduction}

There are some effects that drastically reduce the cross sections
of electron {\selectlanguage{german}bremsstrahlung}
and pair production \cite{Klein}
of atmospheric air showers initiated by high energy astroparticles. 
These suppression
mechanisms can affect air showers by lengthening the showers, and
consequently moving the
position of the shower maximum deeper into the atmosphere.
Two suppression mechanisms have been studied in the present work, namely,
the Landau-Pomeranchuk-Migdal (LPM) 
effect \cite{LP, Migdal} due to the multiple scattering and the dielectric
suppression \cite{Klein} due to the interaction of the bremsstrahlung  
photons with the atomic electrons in the medium through forward
Compton scattering.\\ 
By means of numerical simulations we have studied the influence of these
effects on the development of air showers. We have used the AIRES program
to perform such simulations. AIRES \cite{sergio} represents a set of
programs to simulate atmospheric air showers and to manage all the
associated output data. The physical algorithms of the AIRES system are
based on the realistic procedures of the well-known MOCCA program
\cite{mocca}. AIRES provides some additional features, for example: The
Earth's curvature is taken into account allowing safe operation of all
zenith angles; the simulation programs can be linked to several
alternative
hadronic collission models; etc. To complete our study we have
incorporated new LPM and dielectric suppression algorithms to the AIRES
code.

\section{Migdal theory}

The LPM effect was first predicted by Landau and Pomeranchuk some 40
years ago. Migdal \cite{Migdal} provided the corresponding quantum
mechanical theory
giving analytical expressions for the bremsstrahlung and pair production
cross sections. Recently an experiment performed at SLAC \cite{slac} using
targets of different compositions,
measured the LPM effect founding that there is acceptable agreement
between the experimental data and the Migdal's theory that is presently
considered the standard treatment. \\
Let us
consider first the case of bremsstrahlung where an electron or positron
of energy $E$ emites a photon of energy $k$ in the vecinity of a nucleus
of charge $Z$. The
Migdal cross section for this process is ($c=1$, $\hbar=1$) \cite{Migdal}:

\begin{equation}
\dfrac{d\sigma _{LPM}}{dk}=\frac{4\alpha r_{e}^{2}\xi (s)}{3k}%
\{y^{2}G(s)+2[1+(1-y)^{2}]\phi (s)\}Z^{2}\ln \QOVERD( ) {184}{Z^{\frac{1}{3}%
}}    \label{Blpm}
\end{equation}
where
\begin{equation}
y=\frac{k}{E},    \label{y}
\end{equation}
\begin{equation}
s=\sqrt{\frac{E_{LPM}k}{8E(E-k)\xi (s)}},  \label{s}
\end{equation}
\begin{equation}
G(s)=48s^{2}\left( \frac{\pi }{4}-\frac{1}{2}\int_{0}^{\infty
}e^{-st}\frac{\sin (st)}{\sinh (t/2)}dt\right),   \label{G}
\end{equation}
\begin{equation}
\phi (s)=12s^{2}\left(\int_{0}^{\infty }e^{-st}\coth (t/2)\sin
(st)dt\right) -6\pi s^{2},  \label{phi}
\end{equation}
\begin{equation}
\xi (s)=\left\{ \begin{array}{ll}
             2              & \mbox{if $s <s_{1}$}\\
             1+\ln(s)/\ln(s1) & \mbox{if $s_{1}\leq s \leq 1$}\\
             1              & \mbox{if $s>1$}
\end{array}
\right. 
\label{e}
\end{equation}
($s_{1}= Z^{2/3}/184^{2}$). $r_{e}$ is the classical electron radius 
($r_{e}=e^{2}/m$). $E_{LPM}$ is the characteristic energy of the
LPM
effect and is given by
\begin{equation}
E_{LPM}= \frac{m^{4}X_{0}}{E_{s}^{2}}. \label{Elpm}
\end{equation}
$X_{0}$ is the radiation length and 
\begin{equation}
E_{s}= m\sqrt{4\pi/\alpha}=21.2 \; {\rm MeV}.\label{Es}
\end{equation}
To measure the strengh of the effect it is convenient to introduce the
suppression factor through
\begin{equation}
S=\frac{d\sigma_{LPM}/dk}{d\sigma_{BH}/dk}  \label{S}
\end{equation}  
where $d\sigma_{BH}/dk$ stands for the ``classical'' bremsstrahlung cross
section given by the theory of Bethe and Heitler \cite{BH}.
\latfig{ }{bremprob}{%
\epsfig{file=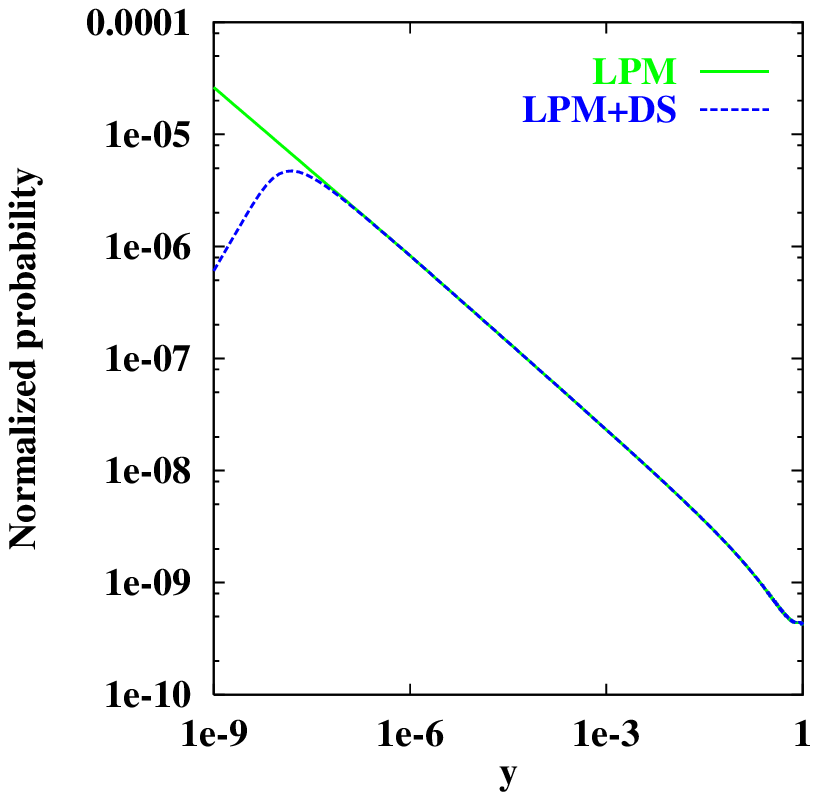,width=8cm}
}{ }{Bremsstrahlung normalized probability versus photon energy, as 
given by Migdal's theory, with and without dielectric suppression. The
electron energy is $10^{18}$ eV and the medium is air (atmospheric depth
$1000\; {\rm g/cm^{2}}$). }

The strenght of the  
effect largely depends on the variable $s$ of equation (3). For $s \ll 1$, 
the suppression is important ($S \ll 1$), while
for $s \gg 1$, there is no
suppression ($S \cong 1$). In fact, when $s\rightarrow \infty $,
the Migdal cross section reproduces, up to 3 \%,  the main terms of the
Bethe-Heitler equation \cite{Klein,BH}.\\
The characteristic energy $E_{LPM}$ gives the energy scale where the
effect
is significant. Notice that $E_{LPM}$ diminishes when the density of the
medium is enlarged. Therefore, for dilute media the LPM effect will be
appreciable only for very high energies. For air in normal conditions, for
example, taking
$Z=7.3$, $\rho=1.2$ ${\rm Kg/m^{3}}$ we have $E_{LPM}=223$ PeV.\\
On the other hand, when $k\rightarrow 0$, it is necessary
to
take into account the change of the photon momemtum due the fact that
the dielectric
constant of the medium is different from one. By considering $k \gg \omega
_{p}$ one obtains: 
$\varepsilon =1-\omega
_{p}^{2}/k^{2}$, where $\omega
_{p}$ is the well-known plasma frequency (for air $\hbar\omega
_{p}=0.73$ eV).
The Migdal approach takes into account this effect usually called
dielectric suppression. 
The influence of the dielectric suppression on the bresstrahlung cross
section is well noted in Figure 1 where the normalized probability for
bresstrahlung is plotted against $y$ in the case of $E=10^{18}$ eV.
Notice how the emission probability is suppressed for $y<10^{-8}$. Since
the energy $E=10^{18}$ eV this corresponds to photon energies $k<0.01$ 
TeV.
\begin{figure}
\begin{displaymath}
\begin{array}{cc}
\hbox{\kern-2mm\epsfig{file=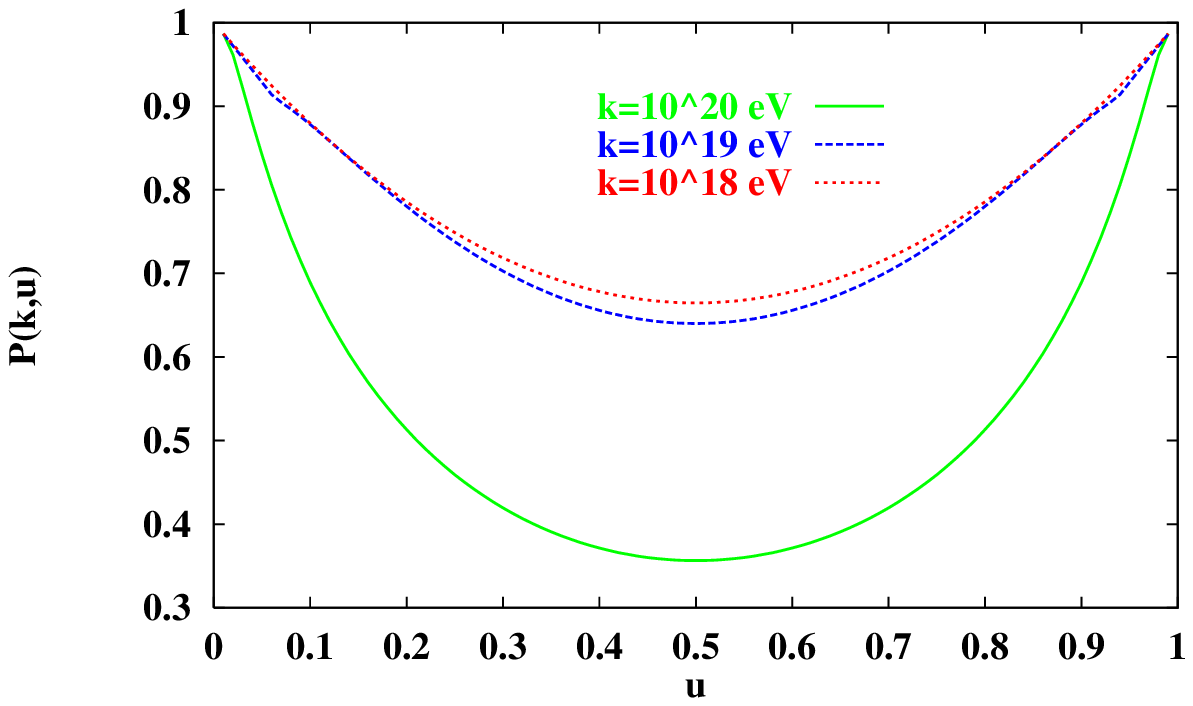,width=7.9cm}} &
\hbox{\kern-0.5cm\epsfig{file=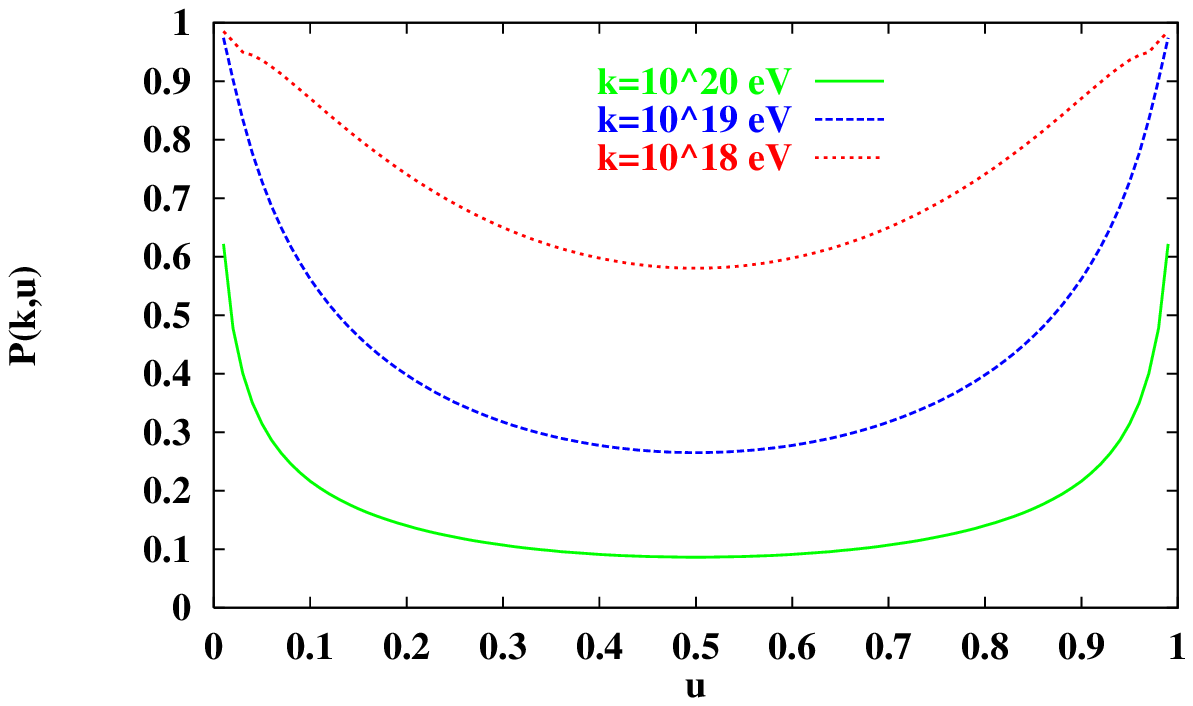,width=7.9cm}\kern-0.5cm}
\end{array}
\end{displaymath}
\caption{ Migdal formulation: Pair production. $a)$  P(k,u), $X=50
g/cm^2$. $b)$ P(k,u), $X=1000 g/cm^2$. }
\end{figure}

We go now into the pair production processes. In this case the cross
section coming from Migdal's
theory reads
\begin{equation}
\dfrac{d\sigma _{LPM}(\gamma \rightarrow e^{+}e^{-})}{dE}=\frac{4\alpha
r_{e}^{2}\xi (\widetilde{s})}{3k}\{G(\widetilde{s})+2[u+(1-u)^{2}]\phi (%
\widetilde{s})\}Z^{2}\ln \QOVERD( ) {184}{Z^{\frac{1}{3}}}  \label{Plpm}
\end{equation}
where
\begin{equation}
u=\frac{E}{k}  \label{u}
\end{equation}
and
\begin{equation}
\widetilde{s}=\sqrt{\frac{E_{LPM}k}{8E(k-E)\xi (\widetilde{s})}}
\label{s-}
\end{equation}
Figure 2 shows the normalized probability of pair production at
atmospheric depth of 50 and 1000 ${\rm g/cm^{2}}$ and for different
energies of
the primary photon.  
Notice how the production probabilities are progresively suppressed when
the primary energy rises. From the figure above, it is also evident that
symmetric processes are
more affectted by the suppression mentioned.
\section{Practical Implementation}

The AIRES simulation system \cite{sergio} has been used as a realistic
program to perform the simulations needed to make our analysis. 
The LPM effect and the dielectric suppression have been incorporated into
the AIRES program.\\
When the characteristics of the atmosphere are taken into account it comes
out that the LPM effect must be considered for all the events where the
energy of the primary is larger than $100$ TeV. This ensures that the
effective cross sections are calculated with a relative error that is
never larger than a few percent.

\section{Simulations}

We have analized the influence of the LPM effect and the dielectric
suppression on the air shower development, performing  simulations for
different primary energies (from
$10^{14}$ to $10^{21}$ eV) and different primary
particles initiating the showers (gamma, proton, electron).
\begin{figure}[b]
\begin{displaymath}
\begin{array}{cc}
\hbox{\epsfig{file=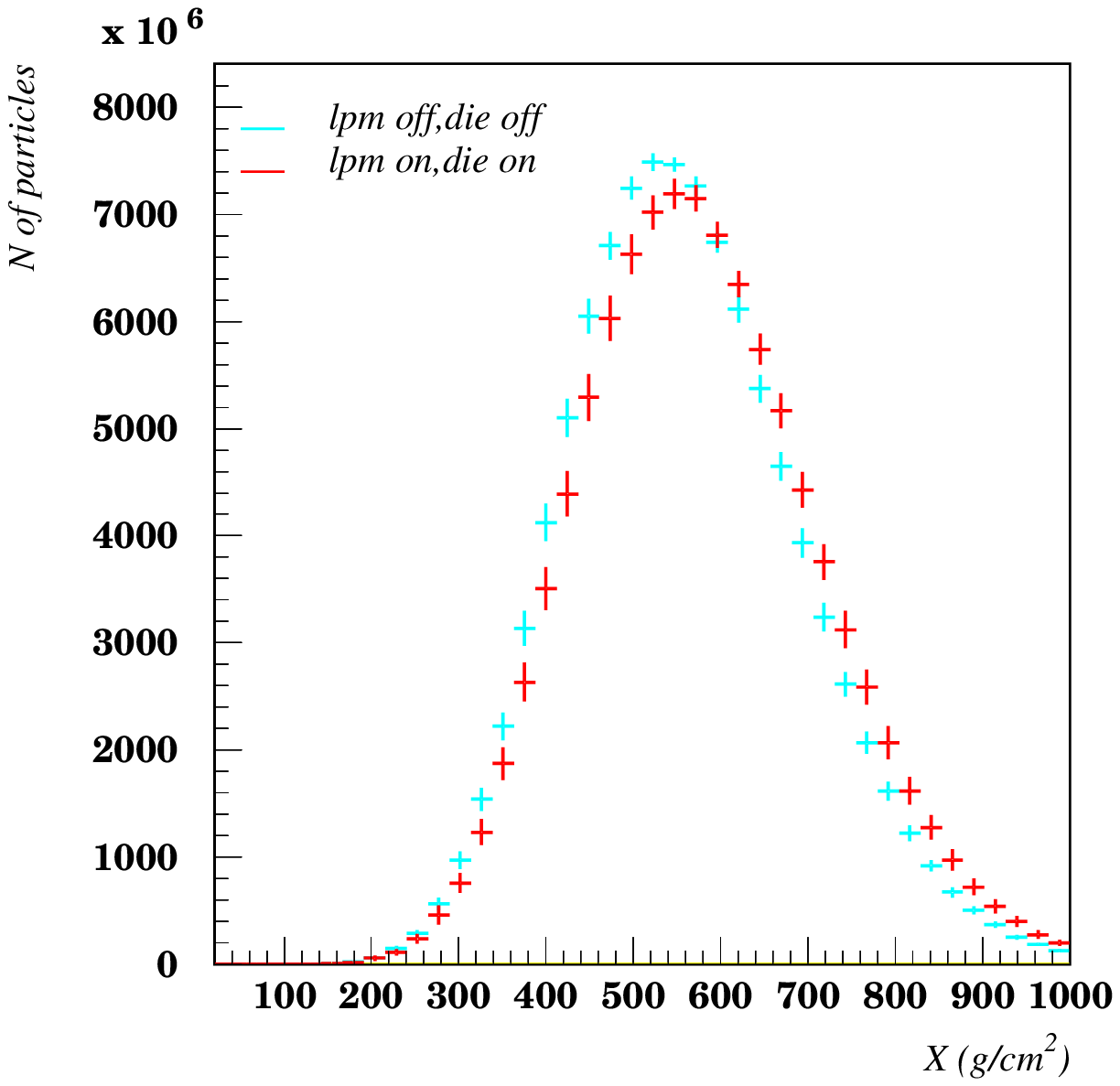,width=7cm}} &
\hbox{\epsfig{file=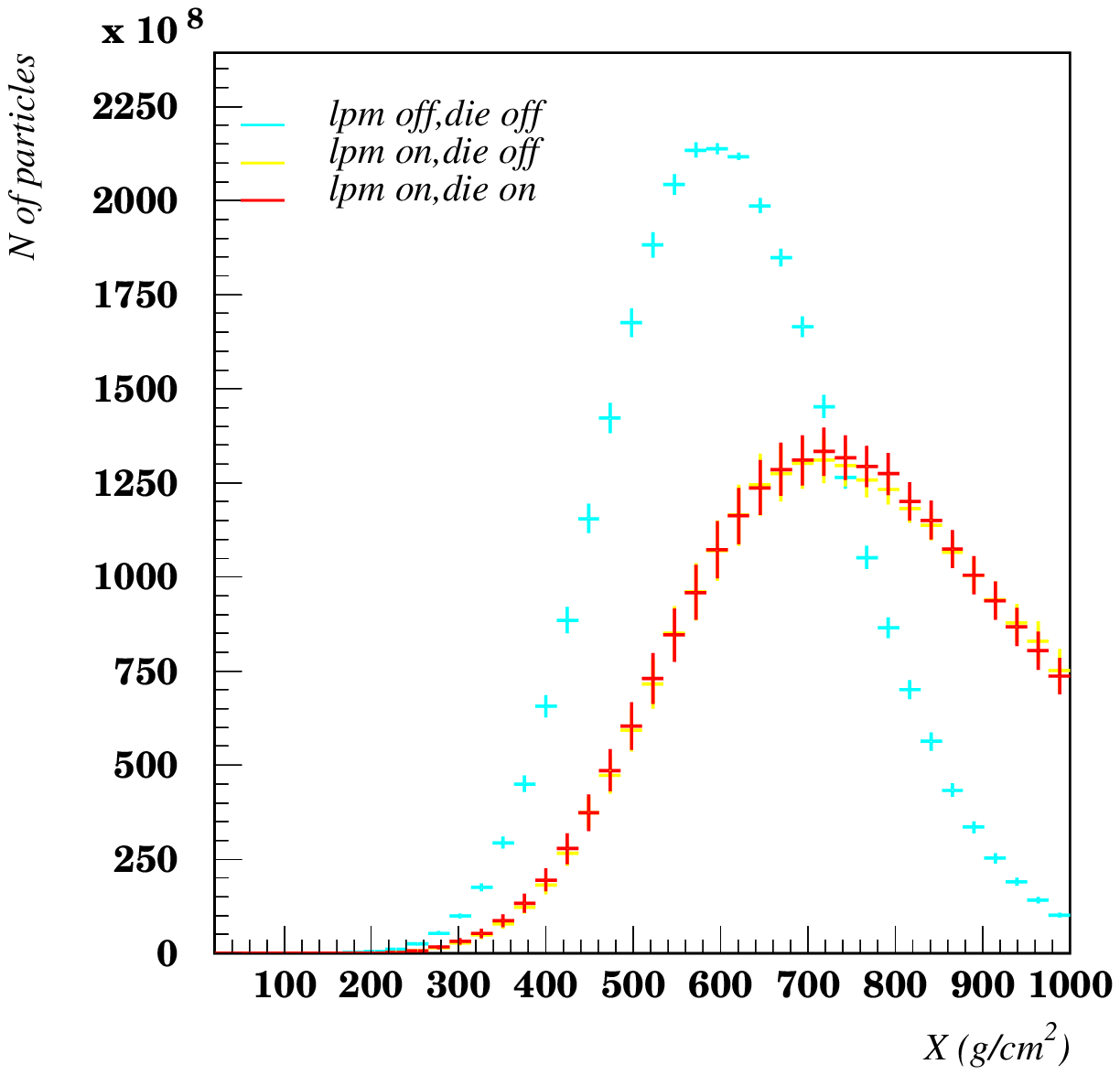,width=7cm}} 
\end{array}
\end{displaymath}   
\caption{Longitudinal development: All charged particles.
Parameters: Primary particle: $\protect\gamma$. Primary energy:
$10^{19}$ eV, $3\times 10^{20}$ eV. Zenith $60^\circ$. Thinning
energy: $10^{-5}$ rel. Injection altitude: $100$ km.}
\end{figure}

For showers initiated by gammas with primary energy larger
than $10^{20}$ the impact of the LPM effect is evident. It affects
the position of $X_{\mathrm{max}}$ (the maximum of the shower), the number
of particles at
$X_{\mathrm{max}}$ and the fluctuations of these magnitudes. 
This is shown in
Figure 3, where the total number of charged particles is plotted against
the vertical depth (longitudinal development of all charged particles)
for two different primary energies. 
The LPM effect affects the gamma showers lengthening and
consequently moving the average position of 
$X_{\mathrm{max}}$ deeper into the atmosphere. This effect is evident in
Figure 4.a where one sees that $X_{\mathrm{max}}$ is shifted in
approximately $100$
${\rm g/cm^2}$ for $10^{20}$ eV  and $500$ ${\rm g/cm^2}$ for $10^{21}$
eV. The
fluctuactions
of $X_{\mathrm{max}}$ also 
increase when the LPM effect is taken into account as can be seen in 
Figure 4.b. 
The average number of charged particles 
at $X_{\mathrm{max}}$ is reduced and its fluctuations are larger
when the LPM effect is introduced
as can be seen in Figure 5 for gamma showers of more
than  $10^{20}$ eV.
We have not found appreciable differences 
for gamma initiated showers with primary energies less than
$10^{18}$ eV. This allow us to conclude that even if the LPM effect must
be taken into account for all particles with energies larger than $100$
TeV, the fraction of such events is statistically significant only for
electromagnetic showers initiated by primaries with energies larger than
$10^{18}$ eV.
\begin{figure}[p]
\begin{displaymath}
\begin{array}{cc}
\hbox{\kern-2mm\epsfig{file=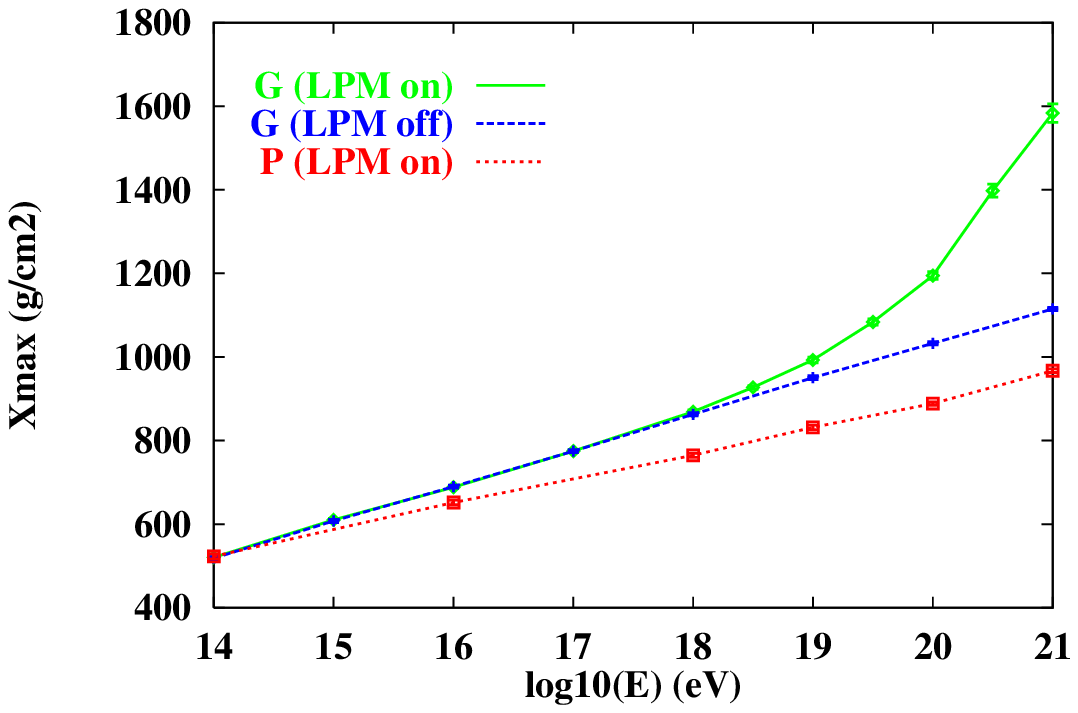,width=7.9cm}} &
\hbox{\kern-5mm\epsfig{file=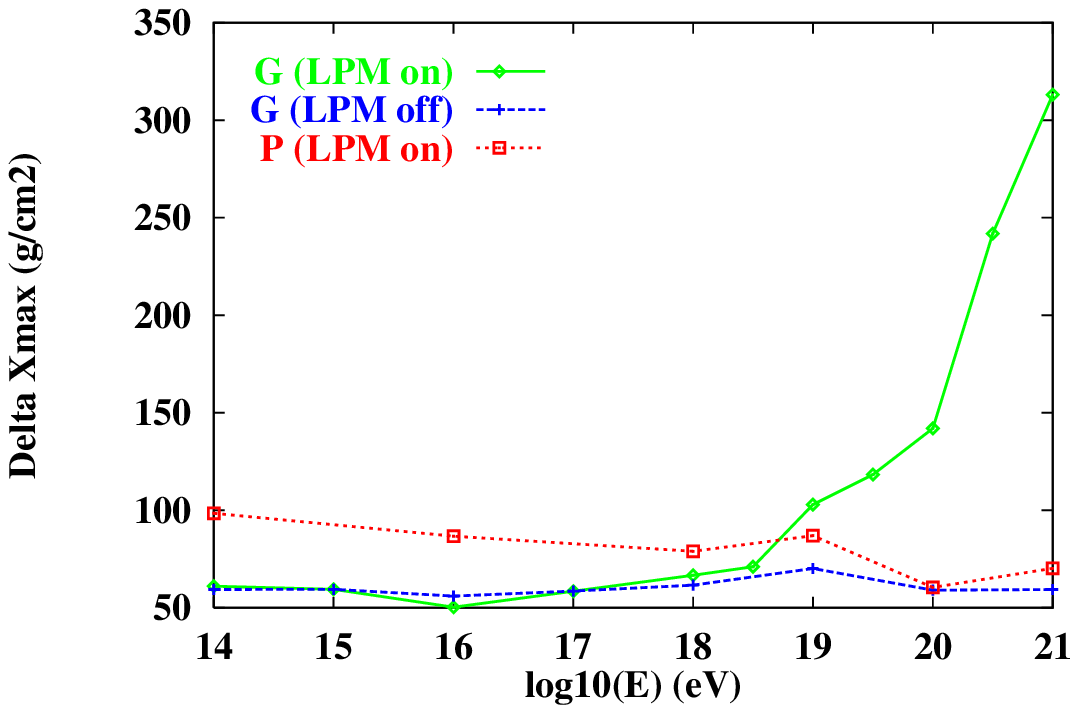,width=7.9cm}\kern-5mm}
\end{array}
\end{displaymath}
\caption{Shower maximum $X_{max}$ (a), and $X_{max}$ fluctuations (b) for 
gamma and proton showers.}
\end{figure}
\begin{figure}[p]
\begin{displaymath}
\begin{array}{cc}
\hbox{\kern-2mm\epsfig{file=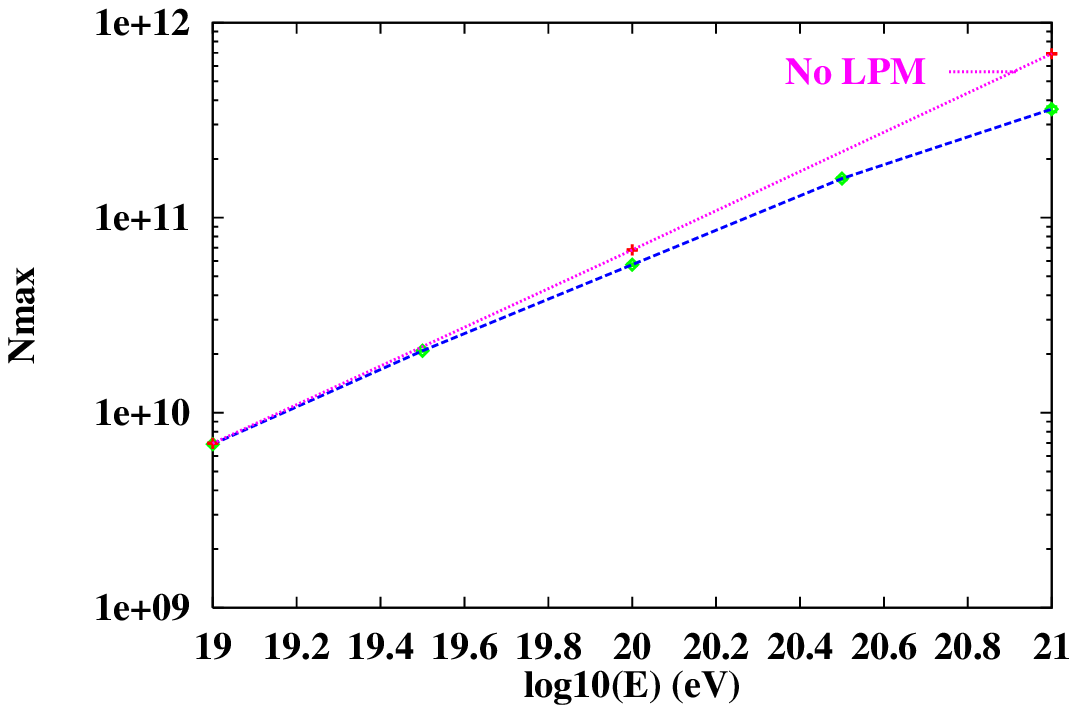,width=7.9cm}} &
\hbox{\kern-5mm\epsfig{file=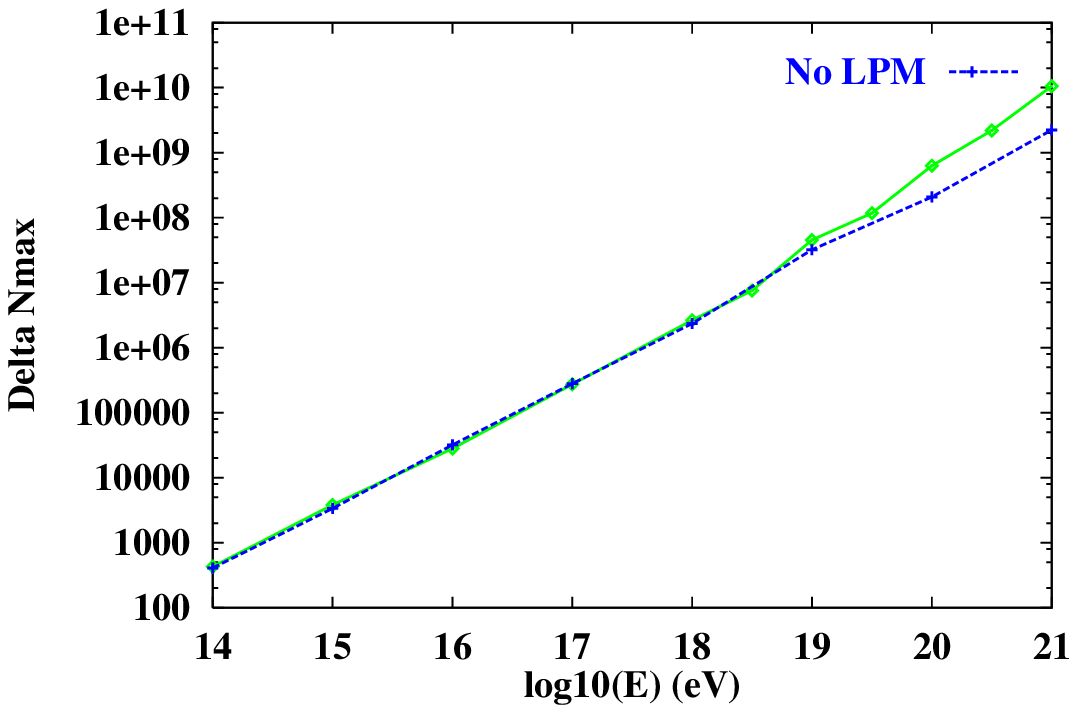,width=7.9cm}\kern-5mm}
\end{array}
\end{displaymath}
\caption{Number of charged particles at $X_{max}$, $N_{max}$ (a), and
$N_{max}$ fluctuations (b) for gamma showers.}
\end{figure}

In Figure 6 the total number of $\mu ^{\pm }$ is plotted
against the vertical depth (longitudinal development of $\mu ^{\pm }$)
againts two different energies.
This observable is very important in the determination of
the
primary composition. In agreement  with the longitudinal development
of all charged particles, it presents the same modifications when the LPM
effect is taken into account.\\
\begin{figure}
\begin{displaymath}
\begin{array}{cc}  
\hbox{\epsfig{file=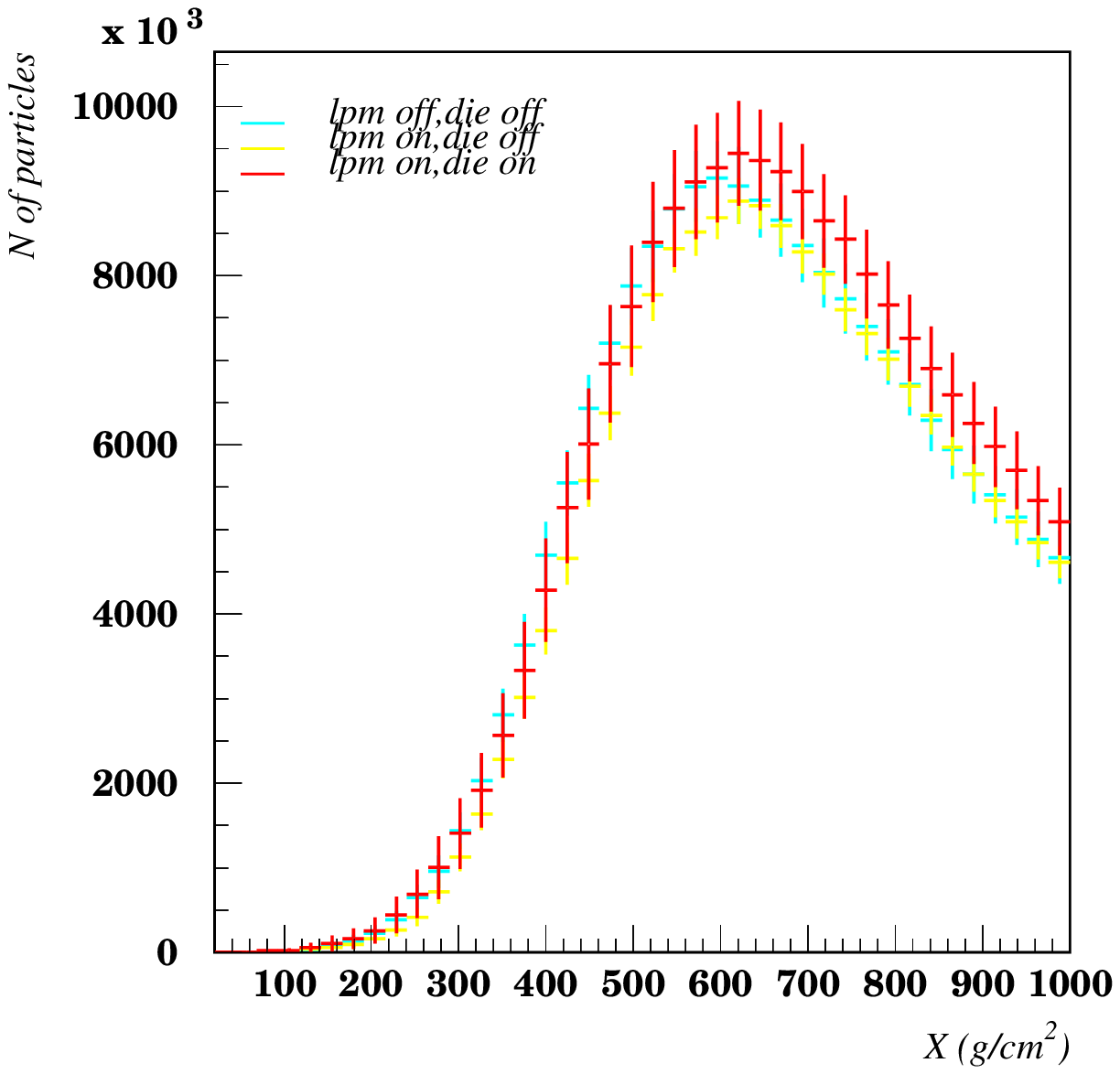,width=7cm}} &
\hbox{\epsfig{file=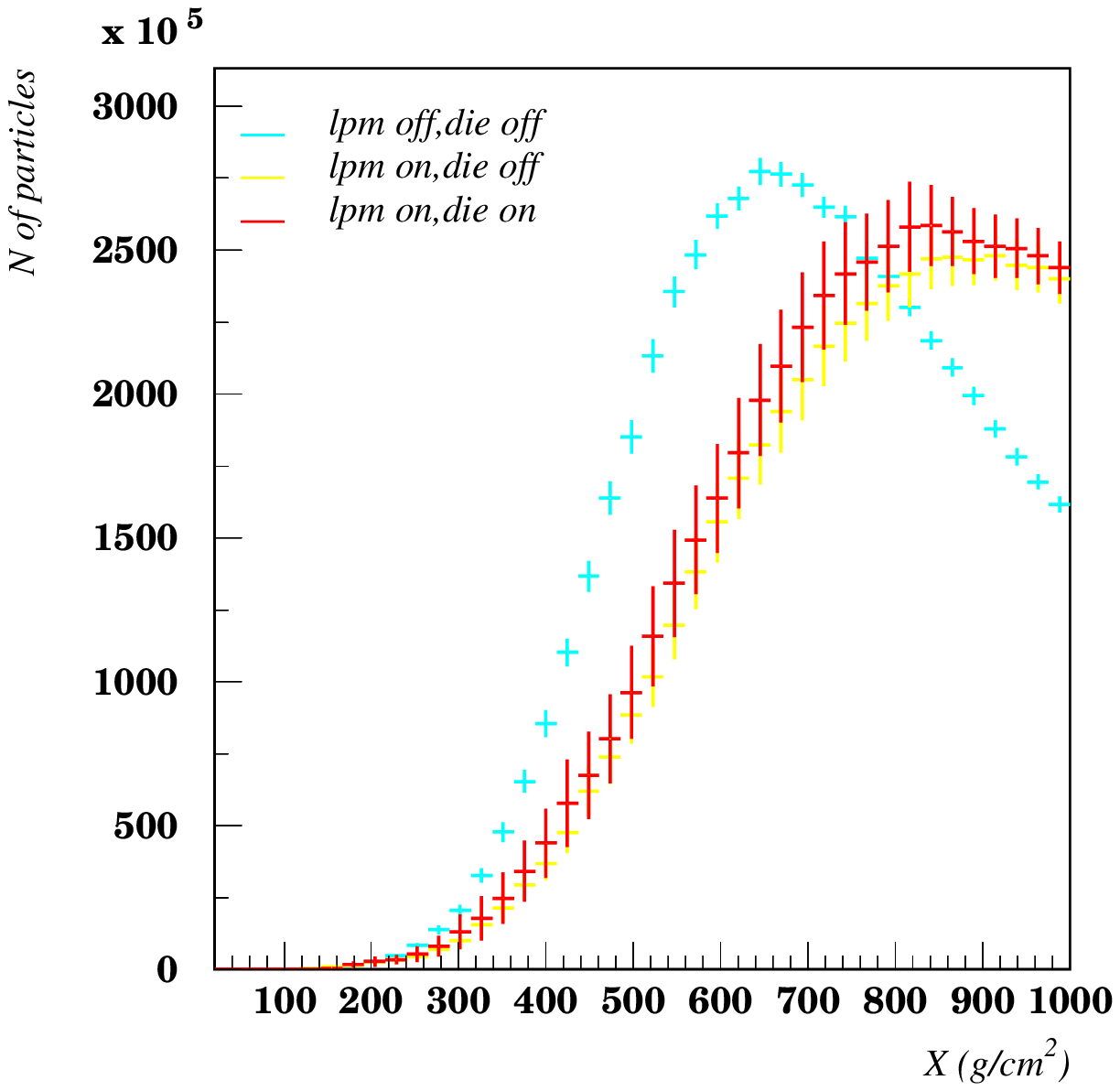,width=7cm}}  
\end{array}
\end{displaymath}
\caption{Longitudinal development of muons.
Parameters: Primary particle: $\protect\gamma$. Primary energy: 
$10^{19}$ eV, $3\times 10^{20}$ eV. Zenith $60^\circ$.
Thinning energy: $10^{-5}$ rel. Injection altitude: 100 km.}
\end{figure}
For air showers initiated by protons, no measurable differences appear
neither
in the average $X_{\mathrm{max}}$, nor 
in the fluctuations of this quantity. The protons interact hadronically,
and
then, the electromagnetic shower, where the LPM effect takes place, starts
later. The proton primary energy is shared among the secondary showers
after
the first interaction and the
electromagnetic cascade begins with energies that are about 2-4 orders
of magnitude less than the
inicial proton
energy. Therefore, one should compare the $10^{20}$ proton
showers with the gamma showers of initial energies of
$10^{17}$-$10^{18}$ eV, where we have not found appreciable differences 
between
showers with and without LPM effect.
The longitudinal development of all charged particles for 
proton initiated showers is plotted in Figure 7.
\begin{figure}
\begin{displaymath}
\begin{array}{cc}
\hbox{\epsfig{file=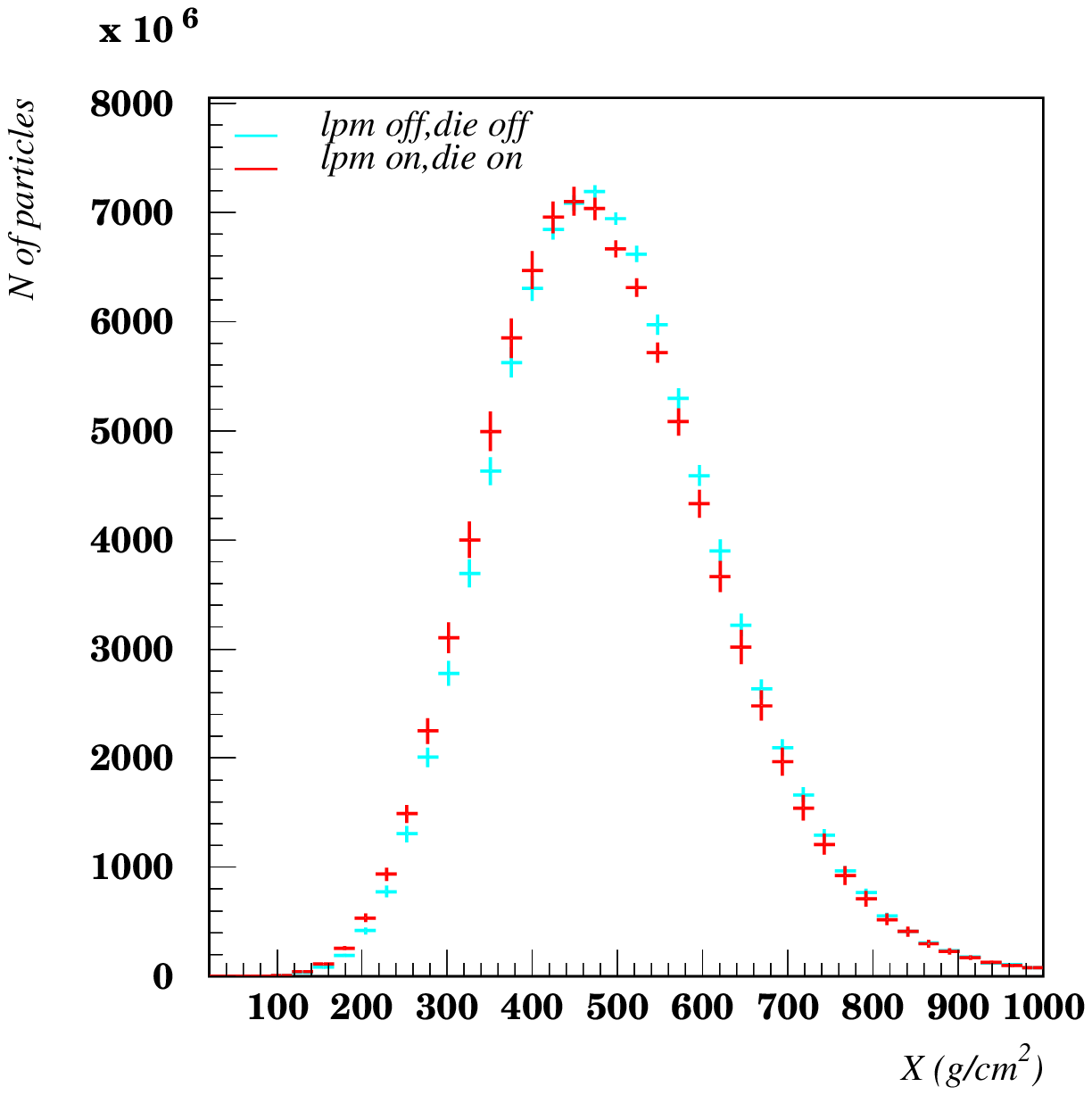,width=7cm}} &
\hbox{\epsfig{file=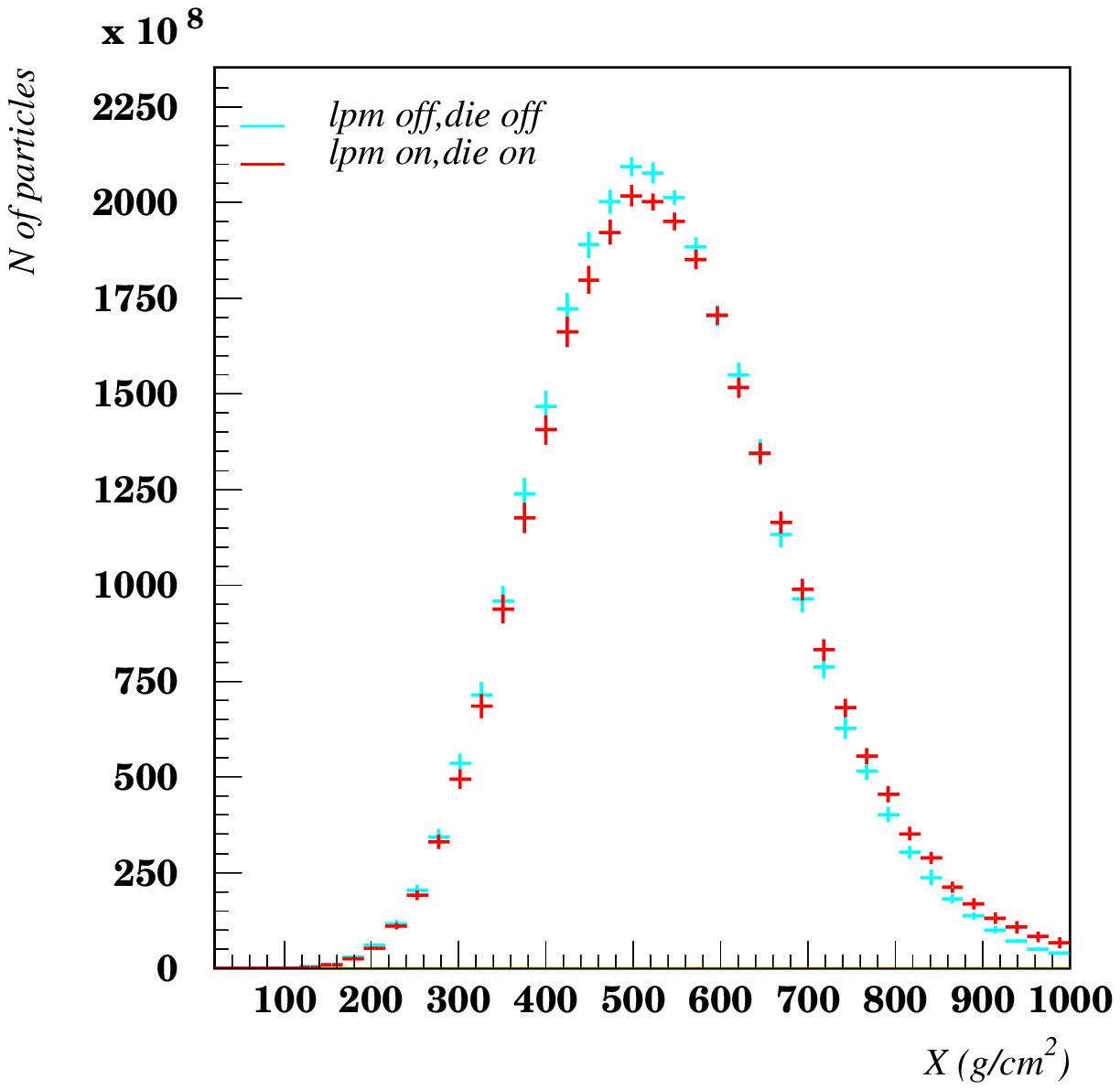,width=7cm}}
\end{array}
\end{displaymath}
\caption{Longitudinal development: All charged particles.
Parameters: Primary particle: $p$. Primary energy:
$10^{19}$ eV, $3\times 10^{20}$ eV. Zenith $60^\circ$. Thinning
energy: $10^{-5}$ rel. Injection altitude: 100 km.}
\end{figure}
%
%
The differences between the simulations with and without
dielectric suppression are less important and are shown in Figure 3 and 6 
for the 
showers initiated by gammas.\\
Finally notice that the characteristics of electron initiated showers
are very similar to those corresponding to gamma showers and for this
reason we have
not included here any related plots.

\section{Conclusions}

The LPM effect introduces modifications on the development of 
gamma
and electron air showers if the
primary energies are larger than $10^{19}$ eV. These effects can be
observed
in the longitudinal development of the showers.
The $X_{\mathrm{max}}$ position for such initial conditions moves
deeper into the atmosphere and its fluctuations are increased when the LPM 
effect is taken into account.
The longitudinal development of $\mu^{\pm }$ also changes in concordance
with all charged particles case.\\
We have not found any significant effect if the showers are initiated by
proton with primary energies up to $10^{21}$ eV, because in this case,
the
electromagnetic shower, where the LPM effect takes place, begins later,
when the initial energy is shared among the secondary particles,
reducing the initial proton energy in 2-4 orders of magnitude. Clearly the
same reasoning is valid for nuclei primary cosmic rays.

\end{document}